\documentclass[useAMS,usenatbib]{mn2e}

\usepackage{graphics}
\usepackage{rotating}
\usepackage{epsfig}
\usepackage{color}

\newlength{\colwidth}
\DeclareGraphicsExtensions{.eps}

\title[Dark matter halo concentrations in WMAP5]
{Dark matter halo concentrations in the {\it Wilkinson Microwave Anisotropy Probe} year 5 cosmology}

\author[A. R. Duffy et al.]
{Alan R. Duffy$^{1,2}$, Joop Schaye$^{2}$, Scott T. Kay$^{1}$ and Claudio Dalla Vecchia$^{2}$  \\
$^1$Jodrell Bank Centre for Astrophysics, Alan Turing Building, The University of Manchester, M13 9PL, U.K.\\
$^2$Leiden Observatory, Leiden University, PO Box 9513, 2300 RA Leiden, The Netherlands}
\begin{document}

\pagerange{\pageref{firstpage}--\pageref{lastpage}} \pubyear{2007}

\maketitle

\label{firstpage}

\begin{abstract}
We use a combination of three large $N$-body simulations to investigate 
the dependence of dark matter halo concentrations on halo mass and
redshift in the {\it Wilkinson Microwave Anisotropy Probe} 
year 5 (WMAP5) cosmology. 
The median relation between concentration and mass is adequately
described by a power-law for halo masses in the range $10^{11} -
10^{15}\,h^{-1}{\rm 
  M}_{\odot}$ and redshifts $z < 2$, regardless of whether the 
halo density profiles are 
fit using NFW or Einasto profiles. Compared with recent
analyses of the Millennium Simulation, which uses a value
of $\sigma_8$ that is higher than allowed by WMAP5, $z=0$ halo
concentrations are reduced by factors ranging 
from 23 per cent at $10^{11}\,h^{-1}{\rm M}_{\odot}$ to 16 per cent at
$10^{14}\,h^{-1}{\rm M}_{\odot}$. The predicted concentrations are
much lower than inferred from X-ray observations of groups and clusters.
\end{abstract}


\section{Introduction}
\label{Introduction}
Investigations into dark matter (hereafter DM) clustering and dynamics
have progressed greatly thanks to large-scale cosmological
simulations run with the aid of supercomputers. In particular, the
$N$-body technique has allowed us to follow the detailed hierarchical
build-up of virialised DM structures, resulting in near spherical
haloes that are well described by the Navarro, Frenk \& White
profile~(\citealt{NFW}, hereafter NFW).  The NFW density profile is
described by just two parameters, the concentration, $c$, and the
total mass, $M$, of the
halo. Simulations have shown that these two parameters are
correlated, with the average concentration of a halo being a weakly
decreasing function of mass
(e.g.\ NFW; \citealt{Bullock:01}; \citealt{ENS:01}, hereafter ENS;
\citealt{Shaw:06,Maccio:07}).
Observations of galaxy 
groups and clusters using X-ray and gravitational lensing data are
being used to test this hypothesis and generally confirm the
anti-correlation between $c$ and $M$
\citep[e.g.][]{Buote:07, SchmidtAllen:07}, although the concentrations
inferred from strong gravitational lensing may exceed those predicted
by the simulations \citep[e.g.][]{Hennawi:07,BroadhurstBarkana:08}.

The best statistical constraints on the halo concentration
distribution currently come from the {\tt Millennium
 Simulation}~\citep{Springel2005}, hereafter {\tt MS}, in which
$2160^3$ DM particles were allowed to interact gravitationally in a
cosmological box of length $500\,h^{-1}{\rm Mpc}$. The resulting
$c(M)$ relation is well described by a power law, $c \propto {\rm M}^{-0.1}$
(\citealt{Neto:07}, hereafter N07). The {\tt MS} used cosmological 
parameters from the {\it Wilkinson Microwave Anisotropy Probe} year 1 (WMAP1)
data release~\citep{WMAP1} combined 
with constraints from the Two Degree Field Galaxy Redshift 
Survey~\citep{Percival}, henceforth referred to as the WMAP1 cosmology.

In this letter, we use a set
of $N$-body simulations to quantify the relation between
concentration and mass in the more recent WMAP year 5
cosmology~\citep{WMAP5}, henceforth WMAP5, in which the most
significant change from WMAP1 is a downward shift in $\sigma_{8}$ 
by around 13 per cent. Because the $c(M)$ relation
is very sensitive to the primordial power spectrum
\cite[e.g.\ ENS;][]{Alam:02,Dolag:04,Kuhlen:05}, it is 
not clear whether the {\tt MS} results accurately describe our
Universe. Indeed, we will show that the WMAP5 cosmology results in a
significant decrease in
both the slope and the normalisation of the $c(M)$ relation. While this
reduction in halo concentrations may make it easier for models of
galaxies to match observations
\cite[e.g.][]{vandenbosch:03,Gnedin:07}, we will show that it
results in strong disagreement between simulations and X-ray observations of
groups and clusters of galaxies.   

\section{Simulations}
\label{Sec:simulations}
We analyse a set of three $N$-body simulations, run using {\sc
 Gadget2}~\citep{Springel2005b}. Each simulation contains $512^3$ DM
particles but with a progressively larger comoving box size: 25, 100, and
$400\,h^{-1}{\rm Mpc}$ for runs {\it L025}, {\it L100} and {\it L400},
respectively. By combining the results of the three simulations, we
cover four orders of magnitude in halo mass, a range that exceeds that
of N07, although our total number of haloes is smaller than in the {\tt MS}. 
Our highest
resolution simulation ({\it L025}) uses a Plummer-equivalent comoving
softening of $2\,{\rm kpc/h}$, with a maximum proper value of
$0.5\,{\rm kpc/h}$, reached at $z=3$; {\it L100} and {\it L400} have
values 4 and 16 times larger respectively. The particle masses are
$8.34 \times 10^{6}\, h^{-1}{\rm M}_{\odot}$, $5.33 \times 10^{8}\,
h^{-1}{\rm M}_{\odot}$ and $3.41 \times 10^{10}\, h^{-1}{\rm
  M}_{\odot}$, for {\it L025}, {\it L100} and {\it L400}, respectively.

Glass-like cosmological initial conditions were generated at redshift
$z=127$ using the Zeldovich approximation and a transfer function
generated using {\sc cmbfast} (v.~4.1, \citealt{CMBFAST}). 
We use the WMAP5 (CMB only) cosmology, with
$[\Omega_{m},$ $\Omega_{b},$ $\Omega_{\Lambda},$ $h,$ $\sigma_{8},$ $n_{\rm s}]$
given by [0.258, 0.0441, 0.742, 0.719, 0.796, 0.963].  For
comparison, the {\tt MS}
used the WMAP1 cosmology, [0.25, 0.045, 0.75, 0.73, 0.9, 1.0];
for which the value of $\sigma_{8}$ is about 13 per cent larger.

\subsection{Halo definitions and density profiles}

Halo virial masses and radii are determined using a spherical 
overdensity routine within the {\sc subfind} algorithm~\citep{subfind}
centred on the main subhalo of Friends-of-Friends (FOF) haloes~\citep{Davis85}.

We perform all calculations for three
different halo definitions, which all take the halo centre to be the
location of the most bound particle in the FOF group. 
According to the first definition, which
is motivated by the spherical top-hat collapse model,
a halo consists of all matter within the radius $r_{\rm vir}$ for which
the mean internal density is $\Delta$ times the critical density
$\rho_{\rm crit}=3H^2/8\pi G$, where $\Delta$ depends on both cosmology and
redshift and is given by \cite{BryanNorman98}. 
Using the second (third) definition, a halo
consists of all matter within the radius $r_{200}$ ($r_{\rm mean}$)
for which the mean internal density is 200 times the critical (mean
background) density. Note that NFW adopted the second definition. We
will use $M_{\rm vir}$, $M_{200}$, and $M_{\rm mean}$ to denote the
corresponding halo masses. 

\begin{figure}
  \begin{center}
    \epsfysize=2in \epsfxsize=4in
    \epsfig{figure=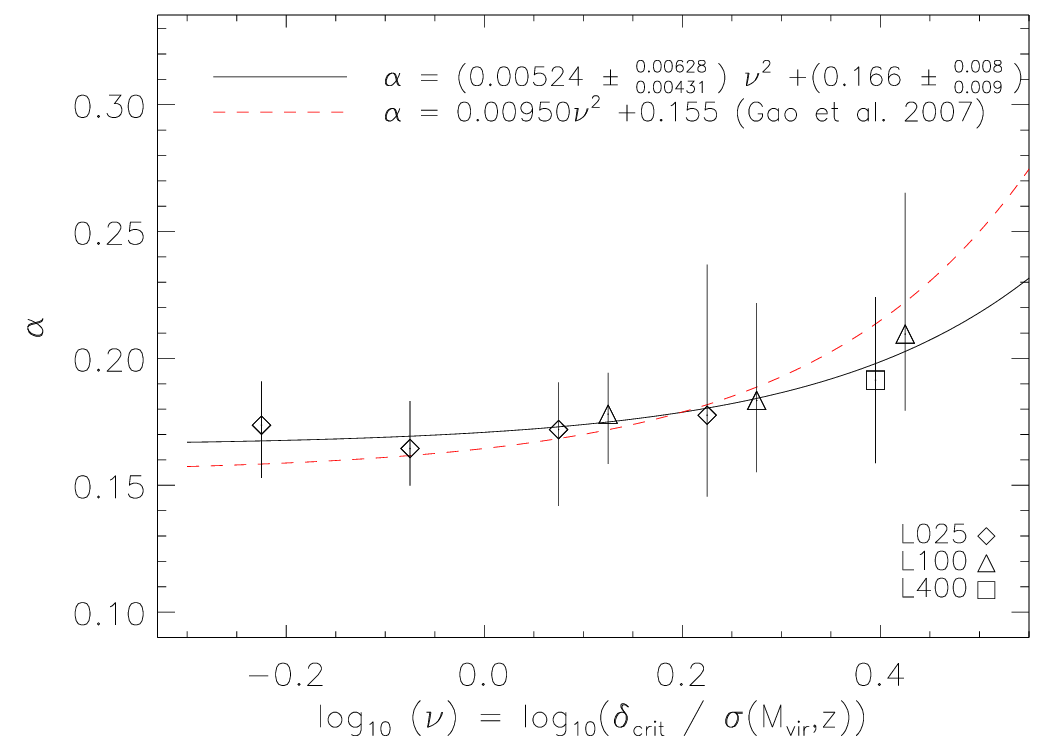, scale=0.45}
    \caption{The relation between the Einasto power-law index
      $\alpha$ and the peak height parameter $\nu$ for stacks of
      haloes of our default sample. Data from  
      redshifts $z=0, 1$ and 2 was combined. Data points correspond to
      medians and error bars indicate quartiles. The 
      solid curve is the best-fit quadratic function, whereas the
      dashed curve is the result from G07. The parameters for both
      fits are given in the legend together with the errors (68.2 per
      cent confidence level) obtained by bootstrap resampling the haloes.}
    \label{fig:einastoalphapeaksimsz}
  \end{center}
\end{figure}

Two different density profiles (NFW and Einasto) are fit to
well-resolved haloes using a procedure which closely follows
N07. For each halo with at least $10^4$ particles within $r_{\rm
  vir}$, all particles in the range $-1.25 \le  
\log_{10} (r/r_{\rm vir}) \le 0$ are binned radially in 
equally spaced logarithmic bins of size $\Delta \log_{10} r =
-0.078$. Density profiles are then fit 
to these bins by minimising the difference between the logarithmic 
densities of the model and the data, assuming equal weighting. 
Haloes are only used if the convergence radius, as proposed 
by~\cite{Power:03}, is within the minimum fit radius of 0.05 $r_{\rm vir}$.
The convergence radius is defined such that the two-body dynamical 
relaxation timescale of the particles internal to this point is 
similar to the age of the universe.
The best-fit density profiles are used to define the halo radii,
with the corresponding masses found by integrating the profiles to 
these radii.

We consider two samples. Our default sample contains all haloes that satisfy 
our resolution criteria while our `relaxed' sample retains only
those haloes for which the separation between the most bound particle
and the centre of mass of the FOF halo is smaller than $0.07 r_{\rm
  vir}$. Note that N07 found that this simple criterion resulted in
the removal of the vast majority of unrelaxed haloes and as such we do
not use their additional criteria. At $z=0$ our
default and relaxed samples contain 1269 and 561 haloes in total.

The NFW density profile is given by
\begin{eqnarray}\label{NFWeqn}
\rho(r) = \rho_{\rm crit} \frac{\delta_{\rm c}}{(r/r_{\rm s})(1+r/r_{\rm s})^{2}}\,,
\end{eqnarray}
where $\delta_{\rm c}$ is a characteristic density contrast and
$r_{\rm s}$ is a scale radius.  The concentration is defined as 
$c_{200} \equiv r_{200}/r_{\rm s}$. NFW profiles were fit
using the two parameters $r_{\rm s}$ and $\delta_{\rm c}$.

The Einasto profile is a rolling power-law first introduced to
describe the distribution of old stars in the Milky Way. It takes the
form
\begin{eqnarray}\label{eqn:Einastoeqn}
\frac{{\rm d} \ln\,\rho}{{\rm d} \ln\,r} = -2
\left(\frac{r}{r_{-2}}\right)^{\alpha}\,,
\end{eqnarray}
where $r_{-2}$ is the radius at which the logarithmic slope of the
density is isothermal (i.e.\ $-2$), analogous to
$r_{\rm s}$ in the NFW profile. As a result the concentration, 
$c_{200}\equiv r_{200}/r_{-2}$, is close to the NFW
value.  Integrating Eq.~\ref{eqn:Einastoeqn} gives
\begin{eqnarray}\label{EinastoInteqn}
\ln(\rho / \rho_{-2}) = -\frac{2}{\alpha}\left[ (r/r_{-2})^{\alpha}
  -1\right]\,,
\end{eqnarray}
where $\rho_{-2}$ is the density at $r_{-2}$. \citet[][hereafter
  G07]{Gao:07} have shown that 
there exists a simple relation between $\alpha$ and $\nu$, the
dimensionless `peak-height' parameter\footnotemark, independent of
redshift. We check this result for our simulations by performing a
three-parameter fit to profiles averaged over ten haloes (to remove
the effects of substructure). As shown in
Fig.~\ref{fig:einastoalphapeaksimsz}, our results are in agreement
with G07; we therefore adopt their $\alpha(\nu)$ relation to reduce
the model to two free parameters ($\rho_{-2}$ and $r_{-2}$).
\footnotetext{The peak height is defined to be $\nu \equiv \delta_{\rm
    crit}/\sigma(M_{\rm vir},z)$, where $\delta_{\rm crit} = 1.686$ is the
  linear density collapse threshold at the present day and
  $\sigma(M_{\rm vir},z)$ is the linear {\it rms} density fluctuation at
  redshift $z$ in a sphere containing a mass $M_{\rm vir}$.}

\section{Results} \label{Sec:concvsmass}

\begin{figure}
  \begin{center}
   \resizebox{\colwidth}{!}{\includegraphics{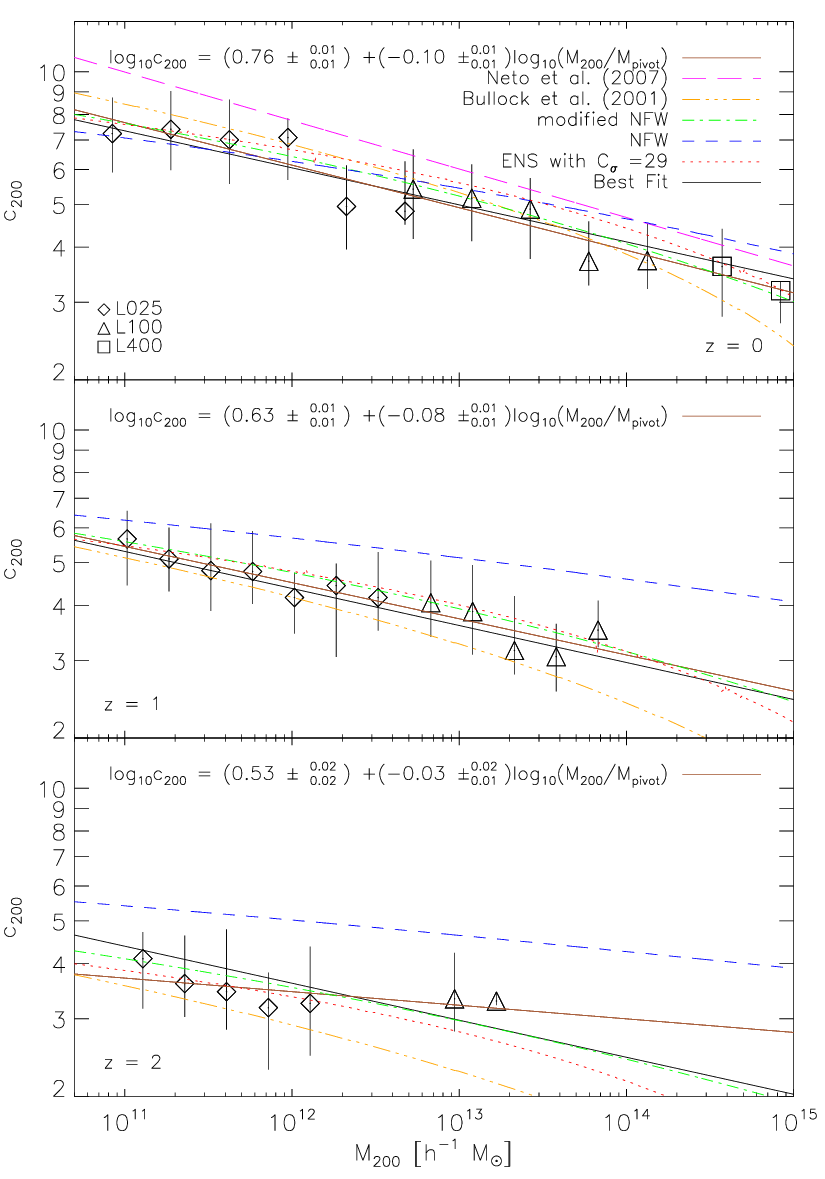}}
    \caption{Concentration-mass relations for $z=0$ (top), 1 (middle)
      and 2 (bottom) using NFW density profiles. Data points
      correspond to median values and error 
      bars to quartiles. Only bins containing at least 5 haloes are
      shown. The black, solid lines show the best-fit power-law
      relation. The errors on the best-fit 
      parameters, given in the legend, are determined by bootstrap 
      resampling the
      haloes and correspond to 68.2 per cent confidence limits. The
      pink dashed line in the top panel shows the best-fit
      power-law relation to all haloes from N07. 
      The other curves represent the
      prescriptions discussed in section~\protect\ref{sec:prescriptions}.
      The brown solid lines correspond tothe special case where we set $C=0$ in 
      Eq. 4, and fit a power-law to the data at each individual redshift.}
    \label{fig:DMconcvsmasssims}
  \end{center}
\end{figure}

Fig.~\ref{fig:DMconcvsmasssims} shows the median $c_{200}(M_{200},z)$
relation for $z=0$, 1 and 2 (top, middle and bottom panels,
respectively) using NFW density profiles.
The data, which spans four orders of magnitude in mass and a
redshift range $z=0 - 2$ are well described by a function of the form
\begin{equation}\label{eq:c(M,z)}
c = A (M/M_{\rm pivot})^B(1+z)^C.
\end{equation}
The black solid line indicates the best-fit relation of this form to the
data. The parameter values and errors are given in Table~\ref{tab:bestfits}
for three different halo definitions, for both NFW and Einasto
profiles, for the full and relaxed samples, and for both $z=0$
(in which case we set $C=0$) and $z=0$--2. All fits use $M_{\rm pivot}
= 2\times 10^{12}\,h^{-1}{\rm M}_{\odot}$, the 
median halo mass, in order to minimise the covariance between $A$ and $B$.  

Concentration is a decreasing function of both mass and redshift,
regardless of the model density profile and halo definition that is
used. 
We note that at high redshift, $z=2$, the mass dependency is
significantly reduced (see also~\citealt{Zhao:03}). 
Compared with NFW concentrations, Einasto concentrations
decrease significantly more rapidly with both mass and redshift. At
$z=0$ the two concentrations agree for $M\sim 10^{14}\,h^{-1}{\rm
  M}_{\odot}$. Concentrations are very sensitive to the halo
definition. In particular, at $z=0$ values for $c_{\rm mean}$ are
nearly twice as large as those for $c_{200}$. As expected, the
difference becomes smaller at high redshift because the critical
density will approach the mean density as the redshift becomes high enough
for matter to dominate over vacuum energy. At $z=0$ the median
concentrations are typically about 10 per cent greater for the relaxed
sample than for the default sample.

Using $C=0$ and $M_{\rm pivot} = 10^{14}\,h^{-1}{\rm
  M}_{\odot}$, N07 found as the best fit
for all NFW haloes in their $z=0$ sample\footnotetext{As a check of our analysis
  procedure, we ran and analysed a miniature version of the {\tt MS}
  with identical mass and force resolution (we used a box of 
  $50\,h^{-1}{\rm Mpc}$ on a side containing $216^3$ particles) and
  were able to reproduce N07's result to within the errors.}
$(A_{200},B_{200}) = (4.67,-0.11)$. Fitting the same function to our
$z=0$ sample gives 
$(A_{200},B_{200}) = (3.93,-0.097)$ which yields concentrations that
are lower by a factor ranging from 23 per cent at $10^{11}\,h^{-1}{\rm
  M}_{\odot}$ to 16 per cent at $10^{14}\,h^{-1}{\rm M}_{\odot}$. This
difference can be attributed to 
the decrease in $\sigma_{8}$.
If  $\sigma_{8}$ is higher then haloes of a given mass form
earlier. This increases the concentration because $c$
reflects the background density of the universe at the time when the
halo forms (NFW).

\begin{table*}
\caption{Best-fit parameters for the median $c(M,z)$ relation
  (Eq.~\protect\ref{eq:c(M,z)}) using  $M_{\rm pivot} = 2\times 
  10^{12}\,h^{-1}{\rm M}_{\odot}$ for three different halo definitions,
  two different density profiles, redshift $z=0$ and $z=0$--2 and for
  both the full (F) and relaxed (R) halo samples. The errors
  correspond to $1\,\sigma$ confidence intervals and have been
  determined by bootstrap resampling the haloes.}
\centering
\begin{tabular}{ccccccccccccc}
\hline 
& & \multicolumn{3}{c}{NFW}  & \multicolumn{3}{c}{Einasto} \\ 
Sample & Redshift & $A_{200}$ & $B_{200}$ & $C_{200}$ & $A_{200}$ & $B_{200}$ & $C_{200}$ \\
\hline 
F & 0 & $5.74 \pm^{0.12}_{0.12}$ & $-0.097 \pm^{0.006}_{0.006}$ & 0 & $6.48 \pm^{0.15}_{0.15}$ & $-0.127 \pm^{0.009}_{0.009}$ & 0 \\
F & 0--2 & $5.71 \pm^{0.12}_{0.12}$ & $-0.084 \pm^{0.006}_{0.006}$ & $-0.47 \pm^{0.04}_{0.04}$ & $6.40 \pm^{0.16}_{0.16}$ & $-0.108 \pm^{0.007}_{0.007}$ & $-0.62 \pm^{0.04}_{0.04}$ \\
R & 0 & $6.67 \pm^{0.11}_{0.11}$ & $-0.092 \pm^{0.007}_{0.007}$ & 0 & $7.70 \pm^{0.14}_{0.15}$ & $-0.127 \pm^{0.008}_{0.009}$ & 0 \\
R & 0--2 & $6.71 \pm^{0.12}_{0.12}$ & $-0.091 \pm^{0.009}_{0.009}$ & $-0.44 \pm^{0.05}_{0.05}$ & $7.74 \pm^{0.15}_{0.16}$ & $-0.123 \pm^{0.009}_{0.008}$ & $-0.60 \pm^{0.05}_{0.05}$ \\
\hline
& & $A_{\rm vir}$ & $B_{\rm vir}$ & $C_{\rm vir}$ & $A_{\rm vir}$ & $B_{\rm vir}$ & $C_{\rm vir}$\\
\hline
F & 0 & $7.96 \pm^{0.17}_{0.17}$ & $-0.091 \pm^{0.007}_{0.007}$ & 0 & $9.03 \pm^{0.23}_{0.23}$ & $-0.122 \pm^{0.008}_{0.009}$ & 0 \\ 
F & 0--2 & $7.85 \pm^{0.17}_{0.18}$ & $-0.081 \pm^{0.006}_{0.006}$ & $-0.71 \pm^{0.04}_{0.04}$ & $8.82 \pm^{0.23}_{0.24}$ & $-0.106 \pm^{0.007}_{0.007}$ & $-0.87 \pm^{0.05}_{0.05}$ \\ 
R & 0 & $9.23 \pm^{0.15}_{0.15}$ & $-0.089 \pm^{0.013}_{0.010}$ & 0 & $10.79 \pm^{0.18}_{0.19}$ & $-0.125 \pm^{0.008}_{0.009}$ & 0 \\ 
R & 0--2 & $9.23 \pm^{0.17}_{0.16}$ & $-0.090 \pm^{0.009}_{0.009}$ & $-0.69 \pm^{0.05}_{0.05}$ & $10.77 \pm^{0.21}_{0.21}$ & $-0.124 \pm^{0.008}_{0.008}$ & $-0.87 \pm^{0.06}_{0.05}$ \\ 
\hline
& & $A_{\rm mean}$ & $B_{\rm mean}$ & $C_{\rm mean}$ & $A_{\rm mean}$ & $B_{\rm mean}$ & $C_{\rm mean}$\\
\hline
F & 0 & $10.39 \pm^{0.22}_{0.22}$ & $-0.089 \pm^{0.006}_{0.007}$ & 0 & $11.84 \pm^{0.29}_{0.30}$ & $-0.124 \pm^{0.007}_{0.008}$ & 0 \\ 
F & 0--2 & $10.14 \pm^{0.22}_{0.23}$ & $-0.081 \pm^{0.006}_{0.006}$ & $-1.01 \pm^{0.04}_{0.04}$ & $11.39 \pm^{0.29}_{0.31}$ & $-0.107 \pm^{0.007}_{0.007}$ & $-1.16 \pm^{0.05}_{0.05}$ \\ 
R & 0 & $12.00 \pm^{0.18}_{0.19}$ & $-0.087 \pm^{0.012}_{0.010}$ & 0 & $14.03 \pm^{0.23}_{0.24}$ & $-0.116 \pm^{0.008}_{0.008}$ & 0 \\ 
R & 0--2 & $11.93 \pm^{0.21}_{0.21}$ & $-0.090 \pm^{0.009}_{0.009}$ & $-0.99 \pm^{0.05}_{0.05}$ & $13.96 \pm^{0.28}_{0.27}$ & $-0.119 \pm^{0.009}_{0.008}$ & $-1.17 \pm^{0.06}_{0.06}$ \\ 
\hline
\end{tabular}
\label{tab:bestfits}
\end{table*} 

\begin{figure}
  \begin{center}
    \epsfysize=2in \epsfxsize=4in
    \epsfig{figure=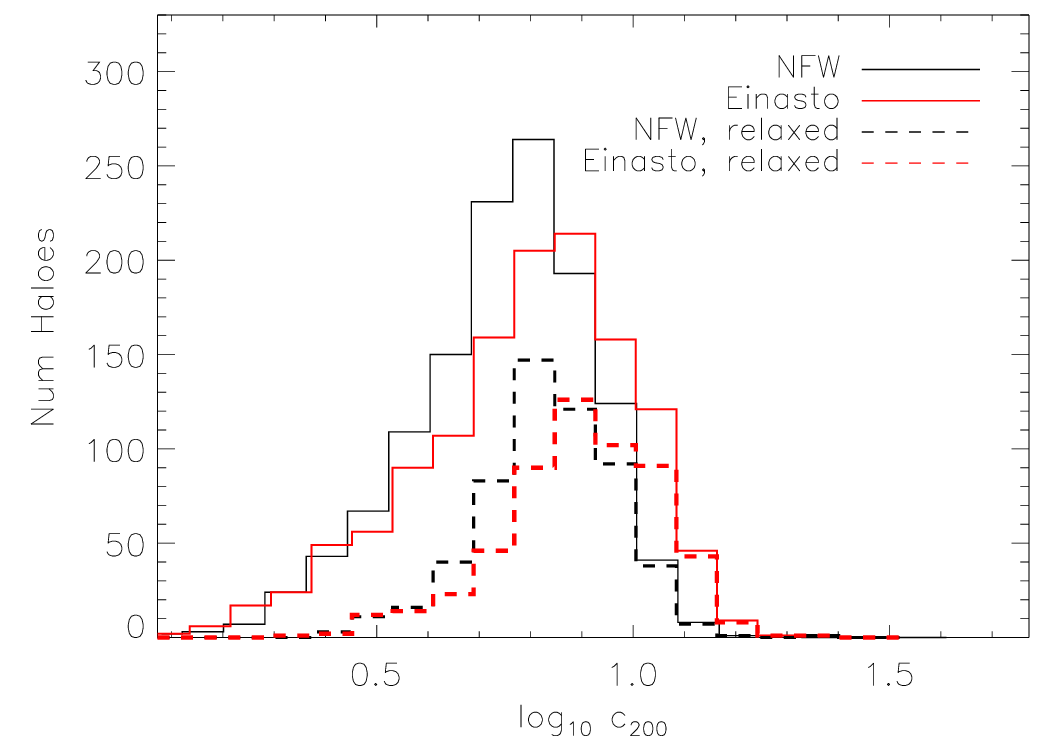, scale= 0.45}
    \caption{Histogram of halo concentrations using the NFW (black) and
      Einasto (red) density profiles for the default (solid) and
      relaxed (dashed) samples.}
    \label{fig:residual}
  \end{center}
\end{figure}

The scatter in the concentration about the median $c(M)$ relation has been
shown to be lognormal for relaxed haloes \citep{Jing:00}, with a slight
decrease in 
the scatter as a function of mass~(N07). The inclusion of unrelaxed
haloes adds a tail towards low concentrations. Fig.~\ref{fig:residual}
shows histograms of the $z=0$ concentrations using both NFW (black) and
Einasto (red) profiles for both the default (solid) and relaxed
(dashed) samples. Either density profile yields a distribution which agrees
qualitatively with that found by N07 for the NFW model. Fitting
lognormal functions to the probability density functions
yields $\sigma(\log_{10} c_{200}) = 0.15$ and $0.17$ for the NFW and
Einasto density profiles, respectively. For the relaxed sample the
scatter is significantly smaller: $\sigma(\log_{10} c_{200}) = 0.11$
and $0.12$. 

\begin{figure}
  \begin{center}
    \epsfysize=2in \epsfxsize=4in
    \epsfig{figure=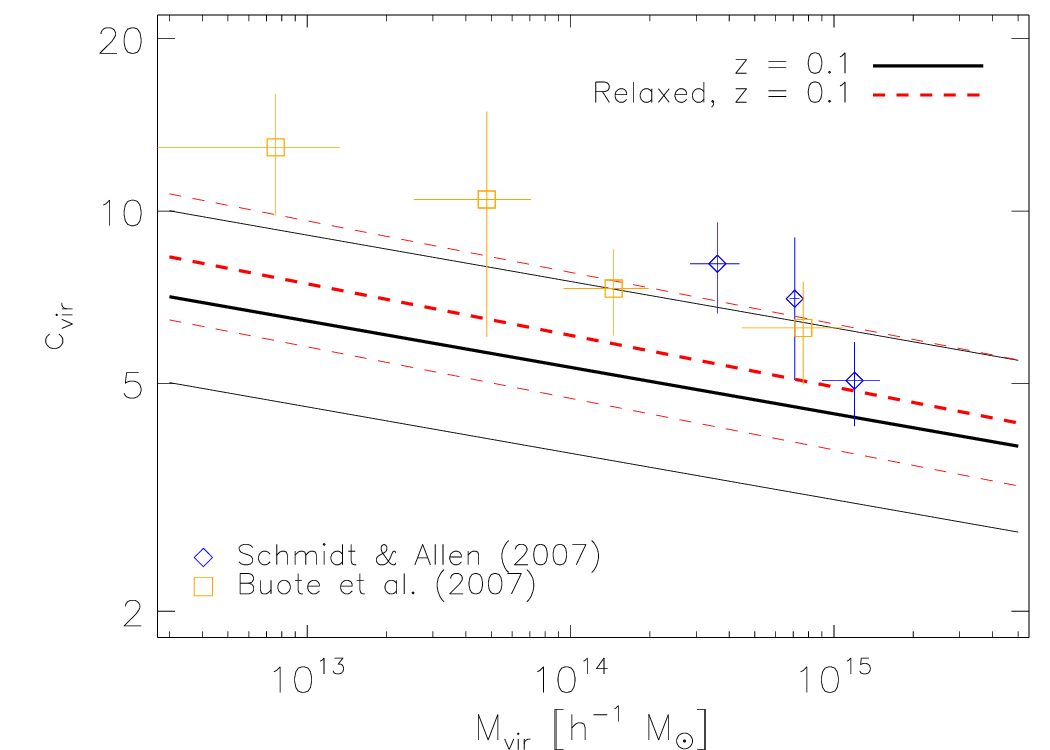, scale= 0.45}
    \caption{Comparison of observed and predicted NFW $c_{\rm vir}(M)$
      relations at $z= 0.1$. Data points indicate weighted averages
      over about 10 objects taken from \citet{Buote:07} (squares)
      and~\citet{SchmidtAllen:07} (diamonds).  
      The sets of solid (dashed) lines are our predictions for the
      median concentrations and the $\pm 1\,\sigma$
      lognormal scatter.} 
    \label{fig:obs}
  \end{center}
\end{figure}

Figure~\ref{fig:obs} compares our predictions with the observations of
\citet{Buote:07} and~\citet{SchmidtAllen:07}, who measured NFW $c_{\rm 
  vir}$ concentrations from X-ray observations of some 70 groups and
clusters with a median redshift of $z=0.1$. The observationally
inferred concentrations are significantly greater than the predicted
medians and the discrepancy increases with decreasing
mass. Using only relaxed haloes (dashed lines) does not help much
because it removes only the low-concentration tail. Observations and
theory could be brought back into agreement if the predicted
concentrations are too low, for example because gas physics has not
been taken into account. Alternatively, the observed sample may be
highly biased towards objects with high concentrations because these
typically have higher X-ray luminosities, particularly for groups.  

\subsection{Comparison with predictions from the literature}
\label{sec:prescriptions}
A number of semi-empirical prescriptions (i.e.\ theoretically motivated
 fitting functions) have been proposed to 
predict the $c(M,z)$ relation for arbitrary cosmologies. The basic
premise of each is that the  
concentration of a halo reflects the background density of the
universe at the formation time of the halo. The prescriptions differ
mainly in the definition of this formation
time. We will compare our results to the prescriptions proposed by
NFW and ENS.\footnote{We also compared with
\citet{Bullock:01}, using the values of $K = 2.9$ and 
$F = 0.001$ \citep{Wechsler:06} and found significant disagreement with
our data at high masses (see Fig.~\ref{fig:DMconcvsmasssims}).} 

The NFW prescription for $c(M)$ has three free parameters. The first two,
$F$ and $f$, are used to define 
the collapse redshift as the redshift for which a fraction $F$ of the
final halo mass $M$ is contained in progenitors of mass $\ge fM$. 
The third parameter, $C$, gives the ratio of $\delta_{\rm c}$, which
is 1-1 related to the concentration parameter in the NFW prescription,
and the mean density of the universe at the collapse redshift. 
The values initially suggested by NFW are $(F,$ $f,$ $C)=(0.5,$
$0.01,$ $3000)$, but we find\footnote{In the original published manuscript we stated these values
incorrectly, this has been updated with a submitted erratum. Values shown here have been corrected.} that the modified values $(0.05$, $0.01,$ $300)$
provide a much better fit to our default sample while the relaxed sample 
is best fit by $(0.23$, $0.01,$ $1000)$ although the latter values are poorly 
constrained. 
Both the original and our modified prescriptions are compared to our data in Fig.~\ref{fig:DMconcvsmasssims}. 
The modified NFW model reproduces the full sample of NFW halo concentrations
over a wide range of masses and redshifts, while the original model strongly
overestimates the concentrations at $z=2$ (bottom panel).

The ENS prescription has only one free parameter, $C_{\sigma}$, which
implicitly defines the collapse redshift through $D(z_c)\sigma_{\rm
  eff}(M_s) = 1/C_\sigma$, where $D(z)$ is the linear growth factor,
$\sigma_{\rm eff}(M)$ is the effective amplitude of 
the linear power spectrum at $z=0$ and $M_{\rm s} = M(r < 2.17
r_{\rm s})$.  The concentration then follows (iteratively) by equating the
characteristic density $\rho_s = 3M/(4\pi r_s^3)$ to the spherical
collapse top-hat density at the 
collapse epoch. ENS found that $C_{\sigma}=28$ was a good fit,
but G07 concluded that this did not describe their data very well (see also~\citealt{Zhao:03}). We
performed a $\chi^{2}$ fit (equally weighting all points) and found
$C_{\sigma}=29$ and 30 as the best-fit values using NFW and Einasto
density profiles, respectively. 
While the ENS prescription is an excellent fit for NFW density profiles below
$z=2$, it underestimates the $z=0$ concentrations for $M< 10^{12}\,h^{-1}{\rm M}_\odot$ when the Einasto density profile is used.

\section{Conclusions}

We have measured the concentrations of DM haloes using
three cosmological $N$-body simulations assuming the WMAP5
cosmology. We presented power-law relations between
halo concentration and mass for $z=0$ and $z=0-2$ 
using both NFW and Einasto density profiles for three different halo
definitions and we compared them to predictions from the
literature. We found that halo concentrations are significantly lower
in the WMAP5 cosmology than in the WMAP1
cosmology, which was used in the {\tt Millennium Simulation}. For $z=0$ the
reduction varies from 23 per cent at $10^{11}\,h^{-1}{\rm M}_{\odot}$
to 16 per cent at $10^{14}\,h^{-1}{\rm M}_{\odot}$. 

While the decrease
in the concentrations may improve the agreement between models and
observations of galaxies, we found that it results in significant 
discrepancies with X-ray observations of groups and clusters of
galaxies. To determine the seriousness of this discrepancy, it will be
necessary to carefully study possible observational selection
biases as well as the effects of baryons on the halo concentrations.  

\section*{Acknowledgements}

We thank V.~Springel for letting us use {\sc subfind} and J.~Bullock, 
C.~Power and especially L.~Gao for helpful discussions. 
We gratefully acknowledge J.~Navarro's help in the use of {\sc
  charden} and {\sc cens}, which we used for the $c(M)$ prescriptions for 
NFW and ENS, respectively. The simulations 
were run on the Cosmology 
Machine at the Institute for Computational Cosmology in Durham as part
of the Virgo Consortium research programme and on Stella, the
LOFAR BlueGene/L system in Groningen. This work was supported by an
STFC studentship, a Marie Curie ETS grant,
an ESF Short Visit grant, and by Marie Curie Excellence Grant
MEXT-CT-2004-014112. 

{\it A special thank you to Anson Daloisio for drawing our attention to a typographical error
in the published version of this manuscript.}

\label{lastpage}
\end{document}